\documentclass{article}

\usepackage{amssymb} % for mathbb
\usepackage{graphicx} % for figures

\begin{document}
%
% paper title
\title{A Mixed-Fractal Model for Network Traffic}

\author{Li Li\footnote{Corresponding author:
li-li@mail.tsinghua.edu.cn}, Yudong Chen, Yi Zhang\\
\\
Department of Automation, Tsinghua University, Beijing, China
100084}% <-this % stops a space
% \newpage is added here to break up from the footnotes

% make the title area
\maketitle

\begin{abstract}
%\boldmath
In this short paper, we propose a new multi-fractal flow model,
aiming to provide a possible explanation for the crossover phenomena
that appear in the estimation of Hurst exponent for network traffic.
It is shown that crossover occurs if the network flow consists of
several components with different Hurst components. Our results
indicate that this model might be useful in network traffic modeling
and simulation.
\end{abstract}

%%%%%%%%%%%%%%%%%%%%%%%%%%%%%%%%%%%%%%%%%%%%%%%%%%%%%%%%%%%%%%%%%%%%
% Section 1.  Introduction
%%%%%%%%%%%%%%%%%%%%%%%%%%%%%%%%%%%%%%%%%%%%%%%%%%%%%%%%%%%%%%%%%%%%
\section{Introduction}
\label{sec:1}

The fractal properties of network traffic were extensively studied
in many literatures during the last decade. It is widely believed
that the dynamic behavior of network flow under re-scaling needs to
be carefully considered in performance analysis and control.

There are numerous explanations and models for the origins and
appearances of fractality for network traffic; see e.g.
\cite{LelandTaqquWillingerWilson1994}-%,
%\cite{ErramilliNarayanWillinger1996},
%\cite{TaqquTeverovskyWillinger1997}, \cite{AbryVeitch1998},
%\cite{AbryFlandrinTaqquVeitch2000}, \cite{MolnarTerdik2001},
%\cite{KaragiannisMolleFaloutsosBroido2004}, \cite{ChenLiZhangHu2009},
%\cite{VeitchHohnAbry2005}, \cite{GongLiuMisraTowsley2005},
\cite{TerdikGyires2009} and the references therein. Under the
multi-fractality assumption for Internet traffic, we propose a new
flow model to explain the crossover phenomena found in the
estimation of Hurst exponent \cite{AbryFlandrinTaqquVeitch2000},
\cite{TerdikGyires2009} (the crossover phenomenon is defined in the
next section). Results indicate that this model help gain more
insights into the actual network traffic dynamics. Moreover, we can
use this model to better simulate network traffic in a simple way.

%%%%%%%%%%%%%%%%%%%%%%%%%%%%%%%%%%%%%%%%%%%%%%%%%%%%%%%%%%%%%%%%%%%%
% Section 2.  Self-Similarity and Multi-Fractality of Network Flow
%%%%%%%%%%%%%%%%%%%%%%%%%%%%%%%%%%%%%%%%%%%%%%%%%%%%%%%%%%%%%%%%%%%%
\section{Self-Similarity and Multi-Fractality of Network Flow}
\label{sec:2}

In data networks, the traffic flow is often viewed a certain
self-similar stochastic process $Y(k)$, which by definition
satisfies \cite{ParkWillinger2000}-\cite{SheluhinSmolskiyOsin2007}
%---------------------------------------------------------------------
\begin{equation}
\label{equ:1} Y(ak) \stackrel{d}{=} a^H Y(k)
\end{equation}
%---------------------------------------------------------------------

\noindent where $a > 0$, and $\stackrel{d}{=}$ denotes the equality
of finite-dimensional distributions. $H > 0$ is the so called Hurst
exponent.

A self-similar process with stationary increments can then be
defined by multiplexing the increments $X(k) = Y(k+1) - Y(k)$ over
non-overlapping blocks of size $n$ as
%---------------------------------------------------------------------
\begin{equation}
\label{equ:2} X^{(n)}(k) = \sum_{j=0}^{n-1} X(kn -j), \indent k \in
\mathbb{Z}
\end{equation}
%---------------------------------------------------------------------

The aggregated process $X^{(n)}(k)$ is called a stationary
self-similar $H$-SSS process with Hurst exponent $H$. It has
finite-dimensional distributions similar to $X(k)$
%---------------------------------------------------------------------
\begin{equation}
\label{equ:3} X^{(n)}(k) \stackrel{d}{=} n^H X (k), \indent n \in
\mathbb{N}
\end{equation}
%---------------------------------------------------------------------

There are various ways to study the stochastic properties of
$X^{(n)}(k)$. \cite{TerdikGyires2009} considers the cumulants of the
aggregated series, which are defined as the Taylor coefficients of
the cumulant-generating function
%---------------------------------------------------------------------
\begin{equation}
\label{equ:5} g(t) = \log \left( E(e^{tX})\right) =
\sum_{m=1}^{\infty} \textrm{cum}_m (X) \frac{t^m}{m!}
\end{equation}
%---------------------------------------------------------------------

\noindent with $\textrm{cum}_m (X) = g^{(m)}(0)$. In
\cite{SamorodnitskiTaqqu1994}, \cite{Terdik1999}, it is shown that
the $m^{\textrm{th}}$ order cumulants of an aggregated $H$-sss
process usually scales as
%---------------------------------------------------------------------
\begin{equation}
\label{equ:4} \textrm{cum}_m \left( X^{(n)} \right)
\stackrel{\Delta}{=} n^{m H(m)} \textrm{cum}_m \left( X(k) \right)
\end{equation}
%---------------------------------------------------------------------

As pointed out in \cite{TerdikGyires2009}, Eq.(\ref{equ:4}) implies
that for each $n, m \in \mathbb{N}$ the logarithm of the modulus of
$\textrm{cum}_m \left( X^{(n)}(k) \right)$ scales linearly with
$\log(n)$ with slope $m H(m)$ as
%---------------------------------------------------------------------
\begin{equation}
\label{equ:6} \log \left( \left| \textrm{cum}_m \left( X^{(n)}
(k)\right) \right| \right) = m H(m) \log(n) + c(m)
\end{equation}
%---------------------------------------------------------------------

The simplest form of $m H(m)$ is a linear function of $m$, i.e.
%---------------------------------------------------------------------
\begin{equation}
\label{equ:7} m H(m) = A m + B.
\end{equation}
%---------------------------------------------------------------------

\noindent In \cite{TerdikGyires2009}, (\ref{equ:7}) is called
linear-fractal model, where the coefficients $A$ and $B$ are
directly estimated (interpolated) during the fittings to the
cumulants.

A special case of the linear-fractal model is the uni-fractal model
\cite{MolnarTerdik2001}\cite{TerdikGyires2009}, which has the form
%---------------------------------------------------------------------
\begin{equation}
\label{equ:8} m H(m) = m + 2 (H - 1)
\end{equation}
%---------------------------------------------------------------------

\noindent where $H$ is the corresponding Hurst exponent.

\cite{TerdikGyires2009} compared the linear-fractal model and the
uni-fractal model using empirical network flow data. As shown in
Fig.3 and Fig.4 of \cite{TerdikGyires2009}, although both models
catch the main trends of the real data in the estimation of Hurst
exponent., neither of them can perfectly match the so-called
crossover phenomenon.

Fig.1 %\ref{fig:1}
gives an illustration of the crossover phenomenon. In the current
case the X-axis stands for $\log(n)$ and the Y-axis stands for
$\log$ cumulant. The slope of the fitting curve crossovers from a
small value to a notably larger value. Therefore the curve consists
of three parts: a line segment with a gentler slope when $\log(n)$
is small, the intermediate transition part, and another line segment
with a steeper slope when $\log(n)$ is large.

%---------------------------------------------------------------------
\begin{figure}[h]
\label{fig:1} \centering
\includegraphics[width=2.8in]{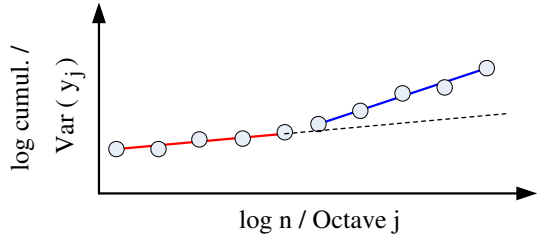}
\caption{An illustration of the crossover phenomena of Hurst
exponent which appears in the fractal model fitting to the cumulants
and the logscale Diagram of wavelet based estimation.}
\end{figure}
%---------------------------------------------------------------------

The crossover phenomenon can also be observed in the wavelet-based
Hurst coefficient estimation for network flow
\cite{AbryFlandrinTaqquVeitch2000}, \cite{VeitchAbry1999}. As shown
in \cite{AbryFlandrinTaqquVeitch2000}, a self-similar process has
the following power law
%---------------------------------------------------------------------
\begin{equation}
\label{equ:9} \textrm{Var} (y_j) = C_y j^{(2 H + 1)}
\end{equation}
%---------------------------------------------------------------------
\noindent where $C_y$ is a constant, $\textrm{octave} ~ j = \log_2
(\textrm{scale})$, and $\textrm{Var} (y_j)$ is a specially defined
measuring variance on $\textrm{octave} ~ j$. Eq.(\ref{equ:9})
indicates that the logarithm of $\textrm{Var} (y_j)$ scales linearly
with $\log (j)$ with slope $(2 H + 1)$ as
%---------------------------------------------------------------------
\begin{equation}
\label{equ:10} \log \textrm{Var} (y_j) = (2 H + 1) \log(j) + \log
C_y
\end{equation}
%---------------------------------------------------------------------

However, real flow data again exhibits the unexpected crossover
phenomenon as illustrated in Fig.1, %\ref{fig:1}
(in this case  the X-axis stands for $\log(j)$ and the Y-axis stands
for $\log$ cumulant). An example can be found in Fig.2.3 of
\cite{AbryFlandrinTaqquVeitch2000}.

\indent

There are numerous physical models for the origin of the
self-similarities in network traffic flows, but few of them can
explain the crossover phenomenon. This gap calls for a finer model
that accounts for the crossover phenomenon, as it will not only
enhance our understanding of the underlying mechanism of the network
traffic, but may also lead to improvements of network performance.

%%%%%%%%%%%%%%%%%%%%%%%%%%%%%%%%%%%%%%%%%%%%%%%%%%%%%%%%%%%%%%%%%%%%
% Section 3.  A Mixed-Fractal Flow Model Yielding Crossover Phenomena
%%%%%%%%%%%%%%%%%%%%%%%%%%%%%%%%%%%%%%%%%%%%%%%%%%%%%%%%%%%%%%%%%%%%
\section{A Mixed-Fractal Flow Model Yielding Crossover Phenomena}
\label{sec:3}

Our previous study \cite{LiHuChenZhang2009} shows that the PCA
eigen-spectrum of the mixed fBm signals with different Hurst
exponents may yield bi-scaling/multi-scaling behavior. Inspired by
that finding, we propose a mixed-fractal flow model, which
reproduces the above crossover phenomena and sheds light on its
origin.

We assume that the network flow process $W(k)$ is the sum of two
independent self-similar processes $X_1(k)$ and $X_2(k)$ with
different Hurst exponents $H_1$ and $H_2$, respectively:
%---------------------------------------------------------------------
\begin{equation}
\label{equ:11} W(k) = \lambda_1 X_1(k) + \lambda_1 X_2(k)
\end{equation}
%---------------------------------------------------------------------

It is assumed that $\textrm{Var}\left\{X_1\right\} =
\textrm{Var}\left\{X_2\right\} = 1$ and thus $\lambda_1$ ,
$\lambda_2>0$ control the variance of the two components. Without
loss of generality, we assume $H_1 < H_2$.

Similarly to Eq. (\ref{equ:2}), we can define $Z^{(n)}(k)$ by
multiplexing the increments $Z(k) = W(k+1) - W(k)$ over
non-overlapping blocks of size $n \in \mathbb{N}$ as
%---------------------------------------------------------------------
\begin{equation}
\label{equ:12} Z^{(n)}(k) = \sum_{j=0}^{n-1} Z(kn -j), \indent k \in
\mathbb{Z}
\end{equation}
%---------------------------------------------------------------------

For the linear-fractal and uni-fratal models, we obtain by
independence
%---------------------------------------------------------------------
\begin{eqnarray}
\label{equ:13} & & \left| \textrm{cum}_m \left( Z^{(n)} (k)\right)
\right| \nonumber \\
& = & \left| \textrm{cum}_m \left( \lambda_1 X_1^{(n)} (k)\right)
\right| + \left| \textrm{cum}_m \left( \lambda_2 X_2^{(n)}
(k)\right) \right| \nonumber \\
& = & c_1(m) n^{m H_1(m)} + c_2(m) n^{m H_2(m)} \nonumber \\
& = & c_1(m) n^{\left[ m + 2(H_1 -1) \right]} + c_2(m) n^{\left[ m +
2(H_2 -1) \right]}
\end{eqnarray}
%---------------------------------------------------------------------

\noindent where $c_1(m)$ and $c_2(m)$ are partly determined by
$\lambda_1$ and $\lambda_2$, respectively. In general, a larger
$\lambda_i$ leads to a larger $c_i(m)$.

If $c_1(m) > c_2(m)$, there is a unique positive solution $n^*$ (not
necessarily an integer) for the following equation in variable $n$
%---------------------------------------------------------------------
\begin{equation}
\label{equ:14} c_1(m) n^{\left[ m + 2(H_1 -1) \right]} = c_2(m)
n^{\left[ m + 2(H_2 -1) \right]}
\end{equation}
%---------------------------------------------------------------------

It is easy to find

%---------------------------------------------------------------------
\begin{equation}
\label{equ:15} \left| \textrm{cum}_m \left( Z^{(n)} (k)\right)
\right| \approx \left\{ \begin{array}{ll}
c_1(m) n^{\left[ m + 2(H_1 -1) \right]} \textrm{, for } n \ll n^* \\
c_2(m) n^{\left[ m + 2(H_2 -1) \right]} \textrm{, for } n \gg n^*
\end{array} \right.
\end{equation}
%---------------------------------------------------------------------

Therefore, for each $m$, the logscale diagram of the
$m^{\textrm{th}}$-order cumulants of $Z^{(n)}(k)$ consist of three
parts: one line segment with slope $m + 2(H_1 - 1)$ when $n$ is
small, the intermediate transition part (which is often short), and
another line segments with slope $m + 2(H_2 - 1)$ when $n$ is large.
This is precisely the crossover phenomenon. An example from
simulation is
given in Fig.2 below. %Fig.\ref{fig:2}

%---------------------------------------------------------------------
\begin{figure}[h]
\label{fig:2} \centering
\includegraphics[width=2.5in]{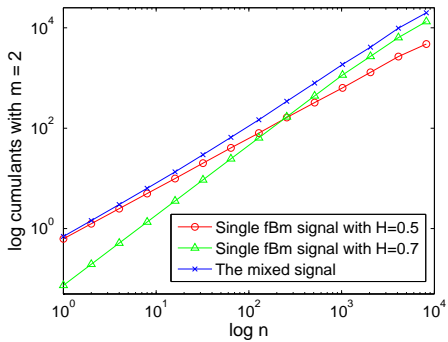}
\caption{The crossover phenomena of Hurst exponent appears in the
cumulant plot, where two self-similar processes with $H_1 = 0.5$ and
$H_2 = 0.7$ are mixed with $\lambda_1 = 2 \lambda_2$. }
\end{figure}
%---------------------------------------------------------------------

Similarly, for the wavelet model, we obtain by independence
%---------------------------------------------------------------------
\begin{equation}
\label{equ:16} \textrm{Var} (z_j) = c_3 j^{(2 H_1 + 1)} + c_4 j^{(2
H_2 + 1)}
\end{equation}
%---------------------------------------------------------------------

\noindent where $c_3$ and $c_4$ are partly determined by $\lambda_1$
and $\lambda_2$, respectively. In general, a larger $\lambda_i$
leads to a larger $c_i$.

If $c_3 > c_4$, there exists a unique positive solution $j^*$ (not
necessarily an integer) for the following equation in variable $j$
%---------------------------------------------------------------------
\begin{equation}
\label{equ:17} c_3 j^{(2 H_1 + 1)} = c_4 j^{(2 H_2 + 1)}
\end{equation}
%---------------------------------------------------------------------

It is easy to find

%---------------------------------------------------------------------
\begin{equation}
\label{equ:18} \textrm{Var} (z_j) \approx \left\{ \begin{array}{ll}
c_3 j^{(2 H_1 + 1)} \textrm{, for } j \ll j^* \\
c_4 j^{(2 H_2 + 1)} \textrm{, for } j \gg j^*
\end{array} \right.
\end{equation}
%---------------------------------------------------------------------

Therefore, the logscale diagram of $Z(k)$ consist of three parts:
one line segment with slope $(2 H_1 + 1)$ when $j$ is small, the
intermediate transition part (which is often short), and another
line segments with slope $(2 H_2 + 1)$ when $j$ is large. An example
from simulation is given in Fig.3. %Fig.\ref{fig:3}

%---------------------------------------------------------------------
\begin{figure}[h]
\label{fig:3} \centering
\includegraphics[width=2.5in]{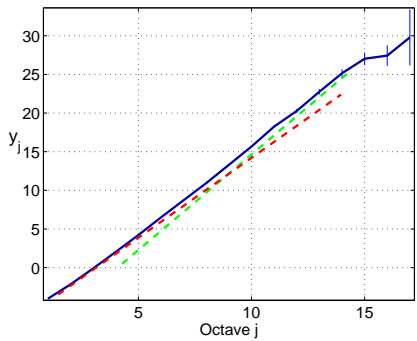}
\caption{The crossover phenomena of Hurst exponent appears in the
logscale Diagram of wavelet based estimation, where two self-similar
processes with $H_1 = 0.5$ and $H_2 = 0.7$ are mixed with $\lambda_1
= 2 \lambda_2$. We apply the Matlab code from \cite{software2007} to
obtain the logscale Diagram here.}
\end{figure}
%---------------------------------------------------------------------

In sum, we have shown that if the traffic flow process consists of
two independent additive components with different Hurst exponents,
we may observe the crossover phenomenon in the estimation of the
Hurst exponents using cumulants or wavelets. It is not hard to see
that this conclusion can be extended to the case with more than two
components.

%%%%%%%%%%%%%%%%%%%%%%%%%%%%%%%%%%%%%%%%%%%%%%%%%%%%%%%%%%%%%%%%%%%%
% Section 4.  Concluding Remarks
%%%%%%%%%%%%%%%%%%%%%%%%%%%%%%%%%%%%%%%%%%%%%%%%%%%%%%%%%%%%%%%%%%%%
\section{Concluding Remarks}
\label{sec:3}

In this section, we would like to index the following points:

First, it is interesting to find that the inconsistence of
theoretical unique-fractal value of the Hurst exponent and the
crossover phenomenon in real data may be resolved by a simple
mixed-fractal model. Moreover, to simulate the mixed-fractal model,
we can generate several processes with different Hurst exponents,
using any of the existing methods for generating uni-fractal
self-similar series, and then add them up. This is in accordance
with the conclusion in \cite{HornaKvalbeinaBlomskoldNilsenb2007}
that we should multiplex multiple flow sources to generate more
appropriate simulated traffic flows.

Secondly, most of the previous physical models for the origins of
self-similarities in network traffic can be adopted to account for
the co-existence of multiple heterogeneous self-similar components.
For example, in the On/Off model
\cite{WillingerTaqquShermanWilson1997}, if it is further assumed
that there exist two different values of the exponent parameters
controlling the Pareto distributions of the ON-period/OFF-period of
sources, then the aggregated flow will have two self-similar
components with different Hurst exponents and thus exhibits
crossover phenomenon. In general, in most complicated data networks,
a channel is shared by many data sources (users) in an approximately
independent and additive way. Due to the diversities of users and
data transfer mechanisms, the inflow from different data sources
might possess different Hurst components. This leads to the
following conjecture: \textit{it is more likely to observe the
crossover phenomenon in the traffic flows on backbone networks .}

Thirdly, using either the cumulant or the wavelet based Hurst
coefficient estimation, we can estimate the approximate varying
range of the Hurst exponents. This work is very helpful, because a
more appropriate and more accurate physical model often allows for a
potential qualitative improvements on network performance
\cite{ParkWillinger2000},
\cite{ParkTuan2000}-\cite{HeGaoHouPark2004}.

% references section

% that's all folks

\begin{thebibliography}{1}

\bibitem{LelandTaqquWillingerWilson1994}
W. E. Leland, M. S. Taqqu, W. Willinger, and D. V. Wilson, ``On the
self-similar nature of Ethernet traffic (extended version),''
\emph{IEEE/ACM Transactions on Networking}, vol. 2, no. 1, pp. 1-5,
Feb. 1994.

\bibitem{ErramilliNarayanWillinger1996}
A. Erramilli, O. Narayan, and W. Willinger, ``Experimental queueing
analysis with long-range dependent packet traffic,'' \emph{IEEE/ACM
Transactions on Networking}, vol. 4, no. 2, pp. 209-223, Apr. 1996.

\bibitem{TaqquTeverovskyWillinger1997}
M. S. Taqqu, V. Teverovsky, and W. Willinger, ``Is network traffic
selfsimilar or multifractal?'' \emph{Fractals}, vol. 5, no. 1, pp.
63-73, 1997.

\bibitem{AbryVeitch1998}
P. Abry and D. Veitch, ``Wavelet analysis of long-range dependent
traffic,'' \emph{IEEE Transactions on Information Theory}, vol. 44,
no. 1, pp. 2-15, Jan. 1998.

\bibitem{AbryFlandrinTaqquVeitch2000}
P. Abry, P. Flandrin, M. S. Taqqu, and D. Veitch, ``Wavelets for the
analysis, estimation and synthesis of scaling data,'' in \emph{Self
Similar Network Traffic Analysis and Performance Evaluation}, K.
Park and W. Willinger, eds., Wiley, 2000.

\bibitem{MolnarTerdik2001}
S. Moln\'{a}r and G. Terdik, ``A general fractal model of Internet
traffic,'' \emph{Proceedings of IEEE Conference on Local Computer
Networks}, Tampa, FL, pp. 492-499, Nov. 2001.

\bibitem{KaragiannisMolleFaloutsosBroido2004}
T. Karagiannis, M. Molle, M. Faloutsos, and A. Broido, ``A
nonstationary Poisson view of Internet traffic,'' \emph{Proceedings
of IEEE INFOCOM}, Hong Kong, vol. 3, pp. 1558-1569, Mar. 2004.

\bibitem{ChenLiZhangHu2009}
Y. Chen, L. Li, Y. Zhang, and J. Hu, ``Fluctuations and pseudo long
range dependence in network flows: A non-stationary Poisson model,''
\emph{Chinese Physics B}, vol. 18, no. 4, pp. 1373-1379, 2009.

\bibitem{VeitchHohnAbry2005}
D. Veitch, N. Hohn, and P. Abry, ``Multifractality in TCP/IP
traffic: the case against,'' \emph{Computer Networks}, vol. 48, no.
3, pp. 293-313, June 2005.

\bibitem{GongLiuMisraTowsley2005}
W.-B. Gong, Y. Liu, V. Misra, and D. Towsley, ``Self-similarity and
long range dependence on the Internet: A second look at the
evidence, origins and implications,'' \emph{Computer Networks}, vol.
48, no. 3, pp. 377-399, June 2005.

\bibitem{TerdikGyires2009}
G. Terdik and T. Gyires, ``L\'{e}vy flights and fractal modeling of
Internet traffic,'' \emph{IEEE/ACM Transactions on Networking}, vol.
17, no. 1, pp. 120-129, Feb. 2009.

%---------------------------------------------------------------------

\bibitem{ParkWillinger2000}
K. Park and W. Willinger, eds., \emph{Self-Similar Network Traffic
and Performance Evaluation}, John Wiley \& Sons, Inc., 2000.

\bibitem{SheluhinSmolskiyOsin2007}
O. Sheluhin, S. Smolskiy, and A. Osin, \emph{Self-Similar Processes
in Telecommunications}, Wiley, 2007.

%---------------------------------------------------------------------

\bibitem{SamorodnitskiTaqqu1994}
G. Samorodnitski and M. S. Taqqu, \emph{Stable Non-Gaussian Random
Processes: Stochastic Models with Infinite Variance}, Chapman \&
Hall/CRC, New York, 1994.

\bibitem{Terdik1999}
G. Terdik, \emph{Bilinear Stochastic Models and Related Problems of
Nonlinear Time Series Analysis: A Frequency Domain Approach},
Lecture Notes in Statistics, vol. 142, Springer Verlag, New York,
1999.

%---------------------------------------------------------------------

\bibitem{VeitchAbry1999}
D. Veitch and P. Abry, ``A wavelet based joint estimator of the
parameters of long-range dependence,'' \emph{IEEE Transactions on
Information Theory}, vol. 45, no. 3, pp. 878-897, April 1999.

\bibitem{LiHuChenZhang2009}
L. Li, J. Hu, Y. Chen, and Y. Zhang, ``PCA based Hurst exponent
estimator for fBm signals under disturbances,'' \emph{IEEE
Transactions on Signal Processing}, in press. (an online pre-print
can be download from
http://ieeexplore.ieee.org/xpl/tocpreprint.jsp?isnumber=4359509\&punumber=78)

\bibitem{software2007}
Matlab code for obtaining the Logscale Diagram is from \\
%http://www.cubinlab.ee.unimelb.edu.au/$\sim$darryl/secondorder\_code.html
%http://www.cubinlab.ee.unimelb.edu.au/$\scriptsize{\sim}$darryl/secondorder\_code.html
http://www.cubinlab.ee.unimelb.edu.au/\~{}darryl/secondorder\_code.html

%---------------------------------------------------------------------

\bibitem{HornaKvalbeinaBlomskoldNilsenb2007}
G. Horna, A. Kvalbeina, J. Blomsk{\o}ld, and E. Nilsenb, ``An
empirical comparison of generators for self similar simulated
traffic,'' \emph{Performance Evaluation}, vol. 64, no. 2, pp.
162-190, Feb. 2007.

\bibitem{WillingerTaqquShermanWilson1997}
W. Willinger, M. S. Taqqu, R. Sherman, D. V. Wilson,
``Self-similarity through high-variability: statistical analysis of
Ethernet LAN traffic at the source level,'' \emph{IEEE/ACM
Transactions on Networking}, vol. 5, no. 1, pp. 71-86, Feb. 1997.

\bibitem{ParkTuan2000}
K. Park, T. Tuan, ``Performance evaluation of multiple time scale
TCP under self-similar traffic conditions,'' \emph{ACM Transactions
on Modeling and Computer}, vol. 10, no. 2, pp. 152-177, April 2000.

\bibitem{HeGaoHouPark2004}
G. He, Y. Gao, J. C. Hou, K. Park, ``A case for exploiting
self-similarity of network traffic in TCP congestion control,''
\emph{Computer Networks}, vol. 45, no. 6, pp. 743-766, Aug. 2004.

\end{thebibliography}
\end{document}